\documentstyle[pre,aps,eqsecnum,epsfig]{revtex}

\begin{document}
\draft
\tightenlines

\title{Elliptic flow in intermediate energy heavy ion collisions and
in-medium effects}

\author{Declan Persram and Charles Gale}

\address{ Physics Department, McGill University\\
         3600 University St., Montr\'eal, QC, H3A 2T8, Canada\\}

\date{\today}

\maketitle
\begin{abstract}
We investigate 
elliptic flow in
heavy ion collisions at intermediate energies. In doing this, we
implement and use a lattice-Hamiltonian model of the nuclear 
interaction and we
also study the effect of in-medium nucleon-nucleon cross sections that follow
consistently from the momentum-dependence of the nuclear mean field.  
\end{abstract}

\section{Introduction }
Heavy ion collisions is an area of research which seeks to study 
physical systems under extreme conditions
of density and temperature in experiments occuring in terrestrial laboratories. 
As such, it represents a rich and challenging field
well worthy of
intellectual pursuit and  its study is important for a deep  
and a more
complete understanding of Nature. At high energies, one goal of this
program is to form and study a new state of matter which is a 
tantalizing prediction
of QCD: the quark-gluon plasma~\cite{qm2001}. At lower energies, the
experimental and theoretical efforts have focused on the need to
characterize and quantify the nuclear equation of state~\cite{nunu97}.
This physics also has an important
role to play in the theory of supernov{\ae} and that of neutron star
properties~\cite{pra97}.  It is that energy regime that we consider
in this work.

 In order to
identify novel many-body features without ambiguity it is
imperative  to provide a realistic model of the nuclear
reaction dynamics. An approach that has proven to be extremely
successful is the Boltzmann-Uehling-Uhlenbeck (BUU) model of heavy
ion collisions~\cite{BDG}. In BUU simulations, nucleons can
suffer hard collisions and can also move on curved trajectories
owing to interactions with the self-consistent mean field. The
interaction term we use in this work is introduced and described in 
Refs.~\cite{GMWelke,JZhang}.
In connection with observables, a considerable amount of information on
the nuclear equation of state and on nonequilibrium precursor phenomena
is accessible through the measurement of global collective behaviour.
A large body of work in intermediate energy
nuclear collisions has been devoted to the measurement and to the
theoretical calculation of both directed 
flow~\cite{PDanielewicz,gut89,eos95,FHaddad2},
and elliptic
flow~\cite{HGutbrod2,YOllitrault,MBTsang}. For 
reviews, see~\cite{NHerrmannReis}, and references therein.

Near the low end of the intermediate energy spectrum, some studies
have put forward the possibility of observing experimental
signatures of new phenomena. A good example is that of modified 
in-medium nucleon-nucleon  cross 
sections~\cite{GFBertsch4,CAOgilvie2,GDWestfall,DKlakow,delaMota,AHombach}.
Those would signal a departure from vacuum properties. 
While the
confirmation of such manifestations would indeed be extremely
interesting, one must keep in mind that such ``new physics''
issues must be addressed with an approach that incorporates all of
the known physics in a computationally tractable model. Our goal
in this paper is to first briefly describe such a model, and then to 
compare its results with experimental data. At the lower 
beam energies considered in this work, the problem of energy
and momentum conservation in transport models is a pressing one.
In addition, the momentum dependence of the nuclear mean field is
an unavoidable feature, both from the point of view of theory 
and from that of
experiment~\cite{BFriedmann,RMafliet,JPJeuhenne-opt,LPCsernai}. With those 
issues in mind, we
use a momentum-dependent lattice Hamiltonian model as a solution of
the BUU transport equation.
This specific model has previously been used in
\cite{DPersram-mrst98}, and provides for excellent energy and momentum
conservation. A similar approach was developed in parallel in
\cite{PDanielewicz4}. 
The momentum-independent lattice Hamiltonian method was first used
in \cite{JLenk}.

Our paper is organized as follows: the next section is a brief
discussion of the
momentum-dependent lattice Hamiltonian model. The following section
introduces the self-consistent in-medium correction to the
free-space nucleon-nucleon cross section as dictated by the
functional momentum dependence that is observed in the nuclear
mean field.
Next, a comparison of model results with compiled
and recent measurements of directed and elliptic flow in 
the range of laboratory
bombarding energy $E_k/A: $25$\rightarrow$800 MeV is performed.
We then summarize and conclude.

\section{The model}
\subsection{Lattice Hamiltonian solution of the mean field}
In this work, we use the BUU transport model which characterizes
the time evolution of a system of nucleons during the course of a
collision of two
heavy ions. The BUU equation reads:
\begin{eqnarray}
\label{BUU-eqn}
\frac{\partial f(\vec{r},\vec{p},t)}{\partial t}
        +\nabla_{\!\!\vec{p}}\,h
        \cdot\nabla_{\!\!\vec{r}}f(\vec{r},\vec{p},t)
        -\nabla_{\!\!\vec{r}}\,h
	\cdot\nabla_{\!\!\vec{p}}f(\vec{r},\vec{p},t)
\nonumber \\
        =
        \int d^3p_1\, d\Omega\,'\,
        \left(
                v_{rel}^*\times\frac{d\sigma^*}{d\Omega\,'}
        \right)
        \left(
                f'f_1'\bar{f}\bar{f_1}
               -ff_1\bar{f'}\bar{f_1'}
        \right).
\end{eqnarray}
The left-hand side of (\ref{BUU-eqn}) accounts for the
time evolution of the semi-classical nucleon phase space distribution
function $f$ due to nucleon transport and mean field effects, and 
the right-hand side accounts for binary nucleon-nucleon collisions
which also modify $f$.
The nucleon single-particle Hamiltonian $h=t+u$,
contains both the single particle kinetic energy and the
mean field single particle potential. The starred quantities in the
collision term represent in-medium values of the relative velocity
and differential nucleon-nucleon scattering cross section
respectively, and $\bar{f}$ accounts for Pauli blocking \cite{BDG}. 

For the nucleon-single particle potential, we adopt the
momentum-dependent parameterization for nuclear matter used in \cite{JZhang},
supplemented with symmetry and Coulomb terms 
\begin{eqnarray}
\label{U_single}
u(\vec{r},\vec{p},\tau_3,t)
	=
        A
          \left(
                \frac{\rho(\vec{r},t)}{\rho_0}
          \right)
        +B
          \left(
                \frac{\rho(\vec{r},t)}{\rho_0}
          \right)^\sigma
                +\frac{2C}{\rho_0}
                \int d^3p'\,
                \frac{
                      f(\vec{r},\vec{p}\,',t)}
                             {1+\left(
                                        \frac{\vec{p}-\vec{p}\,'}
                                             {\Lambda}
                                \right)^2}  
\nonumber \\
	-\tau_3 \frac{2D}{\rho_0}
        \left(\rho_n(\vec{r},t)-\rho_p(\vec{r},t)\right)
	+\frac{(\tau_3+1/2)}{4\pi\epsilon_0}
	 \int d^3r'\,
	 \frac{\rho_p(\vec{r}\,',t)}{|\vec{r}-\vec{r}\,'|}\ .
\end{eqnarray}
This functional is amenable to computational applications and also
incorporates nonequilibrium features \cite{GMWelke,JZhang}. 
The strong interaction part of the mean field is fixed by specifying the
value of the parameters $A, B, \sigma, C$ and $\Lambda$, neglecting the
isospin and Coulomb terms. The parameter
set from \cite{JZhang}, deduced for symmetric nuclear matter,
specifies: $A=-322.0$ MeV, $B=352.5$ MeV,
$\sigma=12/11$, $C=62.75$ MeV and $\Lambda=1.58p_f^0$ which
gives $u(p=0)=-72.4$ MeV, $u(p=p_f)=-51.4$ MeV and
$u(p\rightarrow \infty)=+30.5$ MeV for a zero temperature equilibrium
nuclear matter distribution at saturation density $\rho_0=0.16$
fm$^{-3}$.
The zero temperature 
effective mass at the
Fermi surface and nuclear matter compressibility for this
parameterization 
are $m^*/m=0.67$ and $K=210$ MeV
respectively. The charge-dependent part of the nucleon mean field is taken
into account with the isospin and Coulomb contributions to the
single particle potential, where $\tau_3$ is the third component of
isospin of the nucleon and 
$\rho_n$ and $\rho_p$ are the local neutron and
proton densities, respectively.  For the isospin part of the mean field, 
empirically it is found that $D=34\pm4$ MeV. Furthermore, support for $D$
ranging from $27-40$ MeV is found from various phenomenological
investigations \cite[and Refs. therein]{CHLee}.
We adopt
the value of $D=32$ MeV previously used in the context of 
simulations of heavy ion collisions \cite{MBTsang2,BALi,FSZhang,JYLiu}.
The above parameterization of the mean
field provides a good fit to the nuclear equation of state for both symmetric
nuclear matter and pure neutron matter as calculated
in \cite{AAkmal,AAkmal2} for a only slightly larger value of the nuclear
compressibility. Note however that the exact high density behavior of the symmetry
energy is still an open question that stems partly from the uncertainty in
the high density many-body calculations \cite{pra97}, and partly from the
different theoretical paradigms leading to the equation of state \cite{latpra}. 

The nucleon dynamics are realized through the use of the lattice
Hamiltonian \cite{JLenk} equations of motion, whereby the semi-classical
nucleon phase space 
density for $N_{ens}$ systems of $A$ nucleons
is projected onto a configuration space lattice with a finite
grid constant $\delta x$. At lattice site $\alpha$, the
semi-classical phase space density reads:
\begin{equation}
\label{phase-space}
f_\alpha(\vec{p},t)=\frac{1}{N_{ens}}\sum_i^{A\times N_{ens}}
R(\vec{r}_\alpha-\vec{r}_i(t))\,\delta(\vec{p}-\vec{p}_i(t))
\end{equation}
With this, the lattice
Hamiltonian equations of motion for particle $i$ read:
\begin{equation}
\label{LH-EOM-r}
\frac{\partial \vec{r}_i}{\partial t}=
        \frac{\vec{p}_i(t)}{(p_i(t)^2+m^2)^{1/2}}
        +N_{ens}(\delta x)^3
        \sum_\alpha     R(\vec{r}_\alpha-\vec{r}_i(t))
                        \nabla_{\!\!\vec{p}_i}u_\alpha(\vec{p}_i,\tau_3^i,t)
\end{equation}
\begin{equation}
\label{LH-EOM-p}
\frac{\partial \vec{p}_i}{\partial t}=
        -N_{ens}(\delta x)^3
        \sum_\alpha     u_\alpha(\vec{p}_i,\tau_3^i,t)
                        \nabla_{\!\!\vec{r}_i}R(\vec{r}_\alpha-\vec{r}_i(t)),
\end{equation}
where $u_\alpha(\vec{p}_i,\tau_3^i,t)$ is the discretized form of
(\ref{U_single}), and $R$ is the finite width configuration space
nucleon form factor. In this work, we use $N_{ens}=100$, $\delta
x=1.00$ fm and a
quadratic form factor which provides for good energy and linear
momentum conservation as shown in 
\cite{DPersram}.
In one dimension, the 
form factor is defined as $R(x)=(3\delta
x^2/4 - x^{2})/\delta x^3$ for $0\le x \le \delta x/2$,
$R(x)=(3\delta x /2 - x)^2/2\delta x^3$ for
$\delta x/2< x \le 3\delta x/2$ and $R(x)=0$ for $3\delta x/2<x$.
The normalization factor of $1/N_{ens}$
in (\ref{phase-space}) and $1/\delta x^3$ and $1/2\delta x^3$ in $R$ are chosen such that particle number is
conserved. That is:
\begin{equation}
\label{Norm}
A=(\delta x)^3\sum_\alpha \int d^3p\,  f_\alpha(\vec{p},t),
\end{equation}
at all times $t$. Note that the lattice Hamiltonian equations of
motion (\ref{LH-EOM-r}) and (\ref{LH-EOM-p}) are derived from the
{\em total} Hamiltonian $H$, and not the {\em single particle} Hamiltonian
$h$, as is the case with the often-used test-particle
method~\cite{BDG,JLenk}. This work constitutes the first
application (with Refs.~\cite{DPersram-mrst98} and \cite{DPersram}) 
of the lattice Hamiltonian approach with the interaction of
Eq.~(\ref{U_single}) calculated fully dynamically. See also
Ref.~\cite{lario}.

\subsection{In-medium effects in nucleon collisions}
For the calculation of the Uehling-Uhlenbeck collision term, we
follow the cascade procedure outlined in \cite{BDG} in conjunction
with a Pauli blocking algorithm which distinguishes between protons
and neutrons. However, the in-medium scattering cross section is calculated
such that the functional dependence on momentum that is
observed in the nuclear mean field is respected~\cite{DPersram}.
Fermi's Golden Rule gives an
expression for the total free space nucleon-nucleon cross
section in vacuum:
\begin{equation}
\label{Born-approx}
\sigma_{free}=
        \frac{2\pi V}{\hbar}
        \frac{D_f^{(0)}}{v_{rel}^{(0)}}
        |t_{fi}|^2,
\end{equation}
where $v_{rel}^{(0)}$ and $D_f^{(0)}$ are the relative velocity of the colliding
nucleons and density of final states for the two-nucleon scattering
process respectively. Both of these quantities are expressed as the free space
value.
In matter, we write the total in-medium nucleon-nucleon cross as
\begin{equation}
\label{Born-medm}
\sigma^*
        =
        \frac{2\pi V}{\hbar}
        \frac{D_f^*}{v_{rel}^*}
        |t^*_{fi}|^2,
\end{equation}
where the starred quantities represent in-medium values. Detailed
many-body calculations offer support for $t^*_{fi}\sim t_{fi}$, as
expected for low local densities and/or low energy collisions 
\cite[and Refs. therein]{VRPandharipande-med}.
We follow this line of thought here.
In \cite{VRPandharipande-med}, the
in-medium cross-section is calculated as it is here,
however
a simplified momentum-dependence in the mean field potential was used.
We rewrite the in-medium elastic scattering
cross section as:
\begin{equation}
\label{Sig-med}
\sigma^*
        =
        \frac{v_{rel}^{(0)}}{v_{rel}^{*}}
        \frac{D_f^*}{D_f^{(0)}}
\sigma_{free}.
\end{equation}
\begin{figure}[h!]   
\begin{center}
\includegraphics[width=8.5cm]{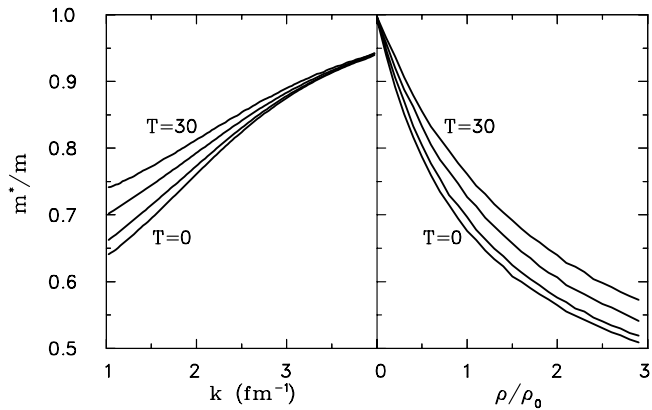}
\end{center}
\caption{Nucleon effective mass in equilibrium nuclear matter as a
function of momentum at saturation density $\rho_0$ (left panel)
and density at the Fermi surface (right panel) for
temperatures of 0, 10, 20 and 30
MeV. In the above, $k=p/\hbar$.}
\label{m_eff-equill-fig}
\end{figure}
We note that a closed form solution to (\ref{Sig-med})
exists for an equilibrium nuclear matter distribution. In this
case, the in-medium nucleon-nucleon total cross section reads:
\begin{equation}
\label{Sigm-med-equill}
\sigma^*
        =
        \left(
        \frac{m^*(\rho,p)}{m}
        \right)^2
\times
\sigma_{free},
\end{equation}
where the nucleon effective mass $m^*$ is defined as:
\begin{equation}
\label{m_eff-equill}
\frac{m^*(\rho,p)}{m}=\left(1+\frac{m}{p}
		\frac{\partial u(\rho,p)}{\partial p}
	\right)^{-1},
\end{equation}
and we have replaced the $\vec{r}$ dependence in $u$ with a $\rho$
dependence which is
appropriate
for an equilibrium
nuclear matter distribution. Note that the density
$\rho(\vec{r}\,)$ is a constant in this case and it enters into the above
relation self-consistently 
via the semi-classical phase space distribution function
$f$ (see, for example, Ref.~\cite{CGale}). In this work a modern and
accurate parameterization of the nucleon-nucleon cross sections is used
\cite{clv}. 
We show in figure \ref{m_eff-equill-fig} the ratio of the nucleon
effective mass to the free mass as a function of momentum,
density and temperature for an equilibrium nuclear matter
distribution.
As seen in the figure, the nucleon effective mass approaches the
free mass in the high momentum, high temperature and low density
limit as is  expected from the many-body nature of this phenomenon. 
\begin{figure}
\begin{center}
\includegraphics[width=7.0cm,angle=270]{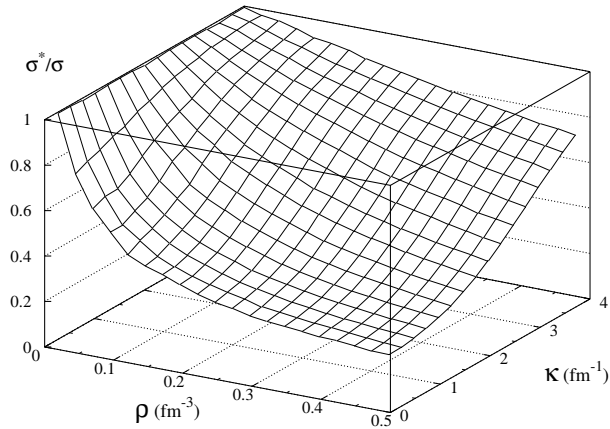}
\end{center}
\caption{Ratio of the in-medium to that of the free space
nucleon-nucleon elastic scattering cross section for an equilibrium
nuclear
matter distribution at $T=0$ MeV.}
\label{sigma-surface-T0}
\end{figure}
In figure \ref{sigma-surface-T0} we show the
ratio of the in-medium to that of the free space nucleon-nucleon
total cross section as a function of both density and momentum for
an equilibrium nuclear matter distribution at $T=0$.
Note that for momentum-independent mean field
potentials, the effective mass is equal to the free mass and
consequently, the in-medium cross section is equal to the free space
cross section.

\subsection{Dynamical effects in heavy ion collisions}
We now wish to apply our model developed thus far to simulate the
collision of two heavy ions. In this scenario, 
dynamical effects will generate a nuclear matter distribution that
is in general not that of a zero temperature system.
Furthermore, one also expects a nuclear matter distribution that is
different from the equilibrium situation. In this case, one cannot
use the zero temperature equilibrium closed form solution of
(\ref{U_single}) as given in \cite{GMWelke}. Instead,
(\ref{U_single}), (\ref{LH-EOM-r}), (\ref{LH-EOM-p}) 
and (\ref{Sig-med}) must 
be explicitly and dynamically calculated at all times during 
the collision process.
In this work, we present the first results
of the self-consistent calculation of (\ref{Sig-med}) as applied to
the dynamical collision of heavy ions.

\begin{figure}[h!]   
\begin{center}
\includegraphics[width=8.5cm]{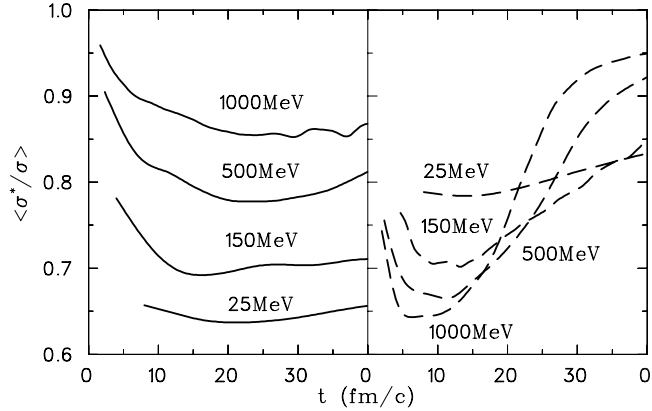}
\end{center}
\caption{Mean value of the ratio of the in-medium to that of the free space
nucleon-nucleon scattering cross section for dynamical collisions of
$^{209}$Bi+$^{209}$Bi at laboratory bombarding energies per nucleon
of 25, 150, 500 and 1000 MeV as a function of time. The left
panel displays the results of the self-consistent calculation of
(\ref{Sig-med}) and the right panel shows the results of the
phenomenological density-dependent ansatz given in
(\ref{Sig-med-density}) with $\alpha=0.20$.}
\label{sigma-rat-E-t}
\end{figure}

\begin{figure}[hb!]   
\begin{center}
\includegraphics[width=8.0cm]{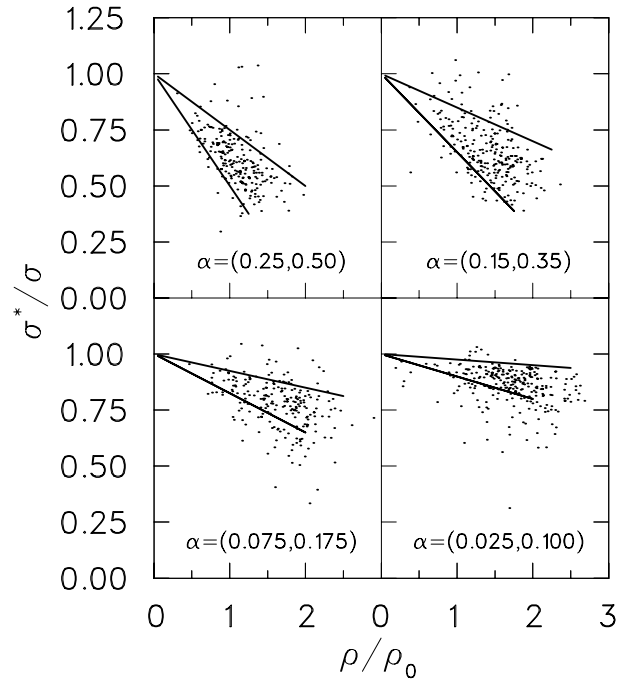}
\end{center}
\caption{Value of the self-consistent in-medium cross section for
the $t:10\rightarrow 20$ fm/c time slice from figure
\protect\ref{sigma-rat-E-t} for all energies considered in that 
figure (scattered points). Moving from left to right and top to bottom,
the panels are for energies of 25, 150, 500 and 1000 MeV.
The solid lines represent the value of the in-medium cross section
using (\protect\ref{Sig-med-density}) with various values of the
parameter $\alpha$.}
\label{sigma-rat-E150}
\end{figure}

In the initial stages of the collision of two heavy ions, large
deviations from equilibrium will result in binary nucleon-nucleon
centre of mass collision energies higher than that at later stages
in the collision process. 
From figure \ref{sigma-surface-T0} one can
infer that initially, owing 
to the density and momentum-dependence in (\ref{Sigm-med-equill}), the
in-medium cross section
will be larger at
earlier stages in the collision compared to later times. In addition,
one can also infer that higher centre of mass nucleus-nucleus
collision energies will result in an in-medium cross section that is
closer to the free space value. We illustrate this for a simulation of
the head-on collision
of two $^{209}$Bi nuclei at laboratory bombarding energies of 25,
150, 500 and 1000 MeV/A in the left panel of figure
\ref{sigma-rat-E-t}. In this figure, we show the mean value of the
in-medium nucleon-nucleon cross section for unblocked collisions as
a function of time. We note that previous attempts to model the
in-medium nucleon-nucleon cross section have used a constant overall scaling
factor \cite{GFBertsch4,CAOgilvie2,Xu}. Others have used a
phenomenological density-dependent in-medium cross section of the
form
\begin{equation}
\label{Sig-med-density}
\sigma^*=(1-\alpha \rho /\rho_0)\sigma_{free},
\end{equation}
where $\alpha$ is an adjustable parameter\cite{GDWestfall,DKlakow}. 
We show in the right
panel of figure \ref{sigma-rat-E-t} that adoption of this simple
ansatz  
results in an in-medium cross section that has considerably
different behaviour from that of the self-consistent calculation. 
The self-consistent modification to the free space cross section in
(\ref{Sig-med}) effectively probes the shape of the
momentum-dependent part of the single particle potential at all
momenta. Such a detailed modeling 
is not possible with neither a constant scaling factor nor
the density-dependent relation in (\ref{Sig-med-density}).
We will
address the consequences of these differences in terms of 
elliptic flow in section
\ref{sec:collective-flow}. Before we leave this section, we show in
figure \ref{sigma-rat-E150} a sample (250 events at each energy) of
the values of the self-consistent in-medium cross section as a
function of density for the $t:10\rightarrow20$ fm/c time
slice from figure \ref{sigma-rat-E-t} for all energies. 
Owing to the additional momentum-dependence in the 
in-medium cross section,
we see that for a given density, the former can take on drastically
different values. The solid lines in that figure indicate the value
of the in-medium cross section that would be obtained by using the
relation given in (\ref{Sig-med-density}) for two different values
of $\alpha$. Again, we see that (\ref{Sig-med-density}) clearly
oversimplifies the in-medium result from (\ref{Sig-med}).

To summarize, we note that the calculation of the self-consistent
momentum (and density) dependent in-medium cross section indicates
that the latter approaches the free space cross section for high
centre of mass nucleus-nucleus collision energies and steadily
decreases with 
centre of mass nucleus-nucleus collision energies. Also, at early
stages in the nucleus-nucleus collision, non-equilibrium effects
results in an in-medium cross section that is closer to the free
space value then compared to later times in the nucleus-nucleus
collision process. This implementation of a self-consistent,
momentum-dependent,  BUU
transport model represents a considerable numerical challenge. Efforts to
develop algorithms in a parallel architecture are under way.

\section{Collective Effects}
\label{sec:collective-flow}
\subsection{Directed Flow}
\label{sec:directed-flow}
\begin{figure}[h!]   
\begin{center}
\includegraphics[width=6.5cm]{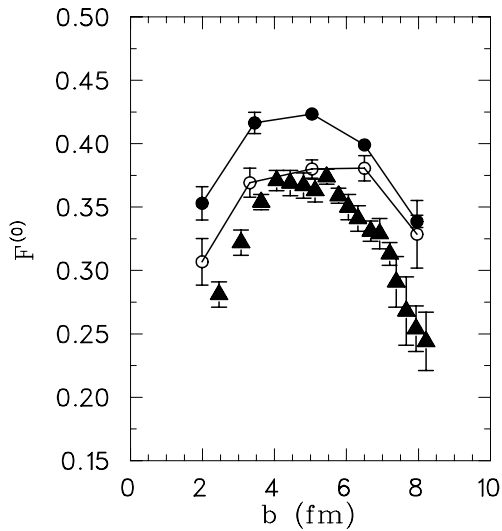}
\end{center}
\caption{Directed flow $F^{(0)}$ as a function of impact parameter for
Au+Au collisions at 400 MeV/A. The measured points
\protect\cite{FRami} shown by triangles, are obtained using a coalescence
model as described in the text. The results from the
BUU calculations are shown by the solid and open circles for a
free space and in-medium cross section from
(\protect\ref{Sig-med}) respectively. The BUU calculations are
shown for all protons at mid-rapidity.
The curves are drawn to guide the eye.}
\label{Flow-b}
\end{figure}

One of the often discussed signatures of collective motion in heavy
ion collisions is the directed flow\cite{PDanielewicz}. 
Typically, the projection into the reaction
plane
of the transverse
momenta of the reaction products is plotted against the rapidity,
resulting in the well known ``S-shaped'' plot. The directed flow
is obtained by taking the slope of this plot at mid-rapidity.
Specifically, one fits the flow plot with:
\begin{equation}
\frac{d<\!p_x\!>}{dy}=a_1+a_2y+a_3y^3,
\label{directed-fit}
\end{equation}
where the rapidity fit interval is taken to be $|y|<y_{proj}$, and
$y_{proj}$ is the initial projectile rapidity. In this case, 
the flow $F$ is given by the value of $a_2$ in (\ref{directed-fit}).
Often, a narrow mid-rapidity region is selected, and (\ref{directed-fit}) is
used to fit the flow, with $a_3=0$. Some recent flow data from the
FOPI collaboration \cite{FRami} is shown in
figure \ref{Flow-b}. 
In this figure, the directed flow for Au+Au
collisions at 400 MeV/A is shown as a function of impact parameter.
The experimental data in 
this plot was obtained by using a coalescence model which weights all
detected fragments by its measured charge\cite{MBTsang}.  We
also show in this figure the results of our BUU simulations with a
free space nucleon-nucleon cross section, and the in-medium cross
section given in (\ref{Sig-med}).  
Here, the  
normalized flow $F^{(0)}=F\times(y_{proj}/p_{proj})_{CM}$ is used.
The coalescence-invariant prescription was developed to enable
direct comparisons with transport models such as the BUU. 
All the
BUU results were obtained by using (\ref{directed-fit}) with and
without $a_3=0$. For the $|y_n|=|y/y_{proj}|<0.45$ cut,
we find the two methods to
be within one standard deviation of each other. Qualitatively, we
find that both the free space and in-medium simulations 
predict the observed maximum in the
flow magnitude for intermediate impact parameters.
More quantitatively, we find the free space cross section over-predicts 
the flow
magnitude for all impact parameters. The
calculations with the in-medium cross section however, are generally in good
agreement with the data at all impact parameters.

\begin{figure}[h!]   
\begin{center}
\includegraphics[width=6.5cm]{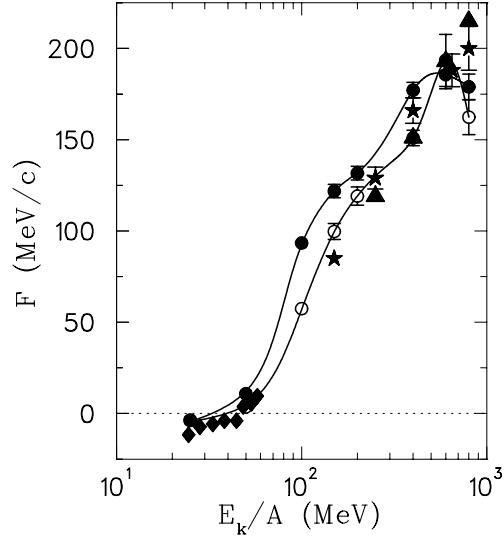}
\end{center}
\caption{Directed flow excitation function for Au+Au collisions. The
measured points \protect\cite{eos95,HAGustafsson}
shown by the triangles and stars are for fragments of charge $Z=1,2$.
The measured points \protect\cite{DJMajestro} shown by the diamonds
are for fragments of
charge $Z=2$. The BUU results (circles)
are as in figure \protect\ref{Flow-b}
and are impact parameter averaged from
$3.5$ fm$<b<6.5$ fm for $E_k/A\ge 100$ MeV at rapidity $-0.19<y_n<+0.29$.
For $E_k/A<100$ MeV, $\hat{b}<0.39$ at rapidity $|y_n|<0.5$, 
fm as is also the case with the measured points shown by the
diamonds. 
The curves are drawn to guide the eye.}
\label{Flow-E-frags}
\end{figure}
The directed flow excitation function also provides a means by which
comparisons of measured signals to model calculations may be
performed. With this in mind, we show in figure \ref{Flow-E-frags} a
compilation of data near the balance energy \cite{DJMajestro} for
central impact parameters and charge $z=2$ fragments
and for semi-central impact parameters at
$E_k/A:150\rightarrow 800$ MeV \cite{eos95,HAGustafsson} for charge
$z=1,2$ fragments. The BUU results for the central impact parameters
are shown for $|y_n|<0.5$ and $\hat{b}=b/b_{max}<0.39$ fm 
and for $-0.19<y_n<+0.29$
and ($3.5<b<6.5$) fm for the semi-central impact parameters.
To compare our BUU results with the measured fragment flow, free
nucleons are omitted from the flow analysis; see Ref. \cite{DPersram}
for a discussion.
\begin{figure}[htb!]   
\begin{center}
\includegraphics[width=6.0cm]{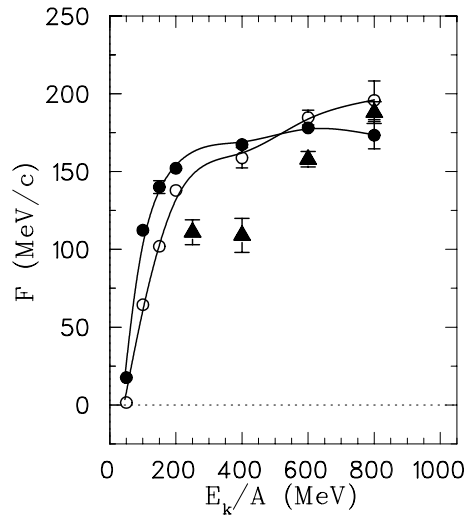}
\end{center}
\caption{Same as figure \protect\ref{Flow-E-frags}, but for 
free protons only.
The measured points \protect\cite{eos95} are shown by triangles and
the BUU results (circles) are as in figure \protect\ref{Flow-b}.
The curves are drawn to guide the eye.}
\label{Flow-E-proton}
\end{figure}
For the
entire energy range probed here, we find that the calculations employing
the in-medium nucleon-nucleon cross section result in a reduced value
of the directed flow as compared with the free space cross section.
From this figure, we find the in-medium result to be in  
better agreement with the data. A more stringent test of the model in
terms of the directed flow is shown in figure \ref{Flow-E-proton}. 
In
this figure, the data for free protons \cite{eos95} is shown along with
the model predictions for free protons. 
The figure indicates that the predicted directed flow over-estimates
the observed value. We point out however, that the 
coalescence-invariant comparison in figure \ref{Flow-b} with the 
more recent data is in better agreement with the model calculations. 

\subsection{Elliptic Flow}
\label{sec:elliptic-flow}
We now turn to comparisons of the measured elliptic flow with our
BUU model. The elliptic flow is a measure that quantifies the azimuthal
anisotropy of the momentum distribution. 
Specifically, we fit the azimuthal distribution of
nucleons about the reaction plane with a Fourier expansion of the
form:
\begin{equation}
\frac{dN}{d\phi}=p_0\left(
	1
	+2p_1cos(\phi)
	+2p_2cos(2\phi)
		\right),
\label{Fourier-fit}
\end{equation}
where the ellipticity coefficient
$p_2$ depends on the in-plane and out-of-plane flow amplitudes. In addition,
the anisotropy ratio is defined by: $R_n=(1- 2 p_2)/(1+ 2 p_2)$. A ratio $R_n$
larger than unity signals a preferred out-of-plane emission. 
\begin{figure}[hb!]   
\begin{center}
\includegraphics[width=7cm]{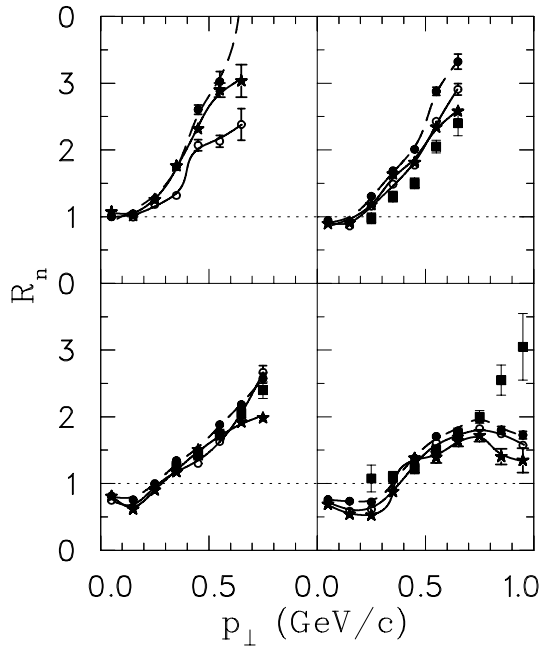}
\end{center}
\caption{Mid-rapidity ($|y_n|<0.15$)
free-proton anisotropy ratio for Bi+Bi collisions at
incident laboratory bombarding energies of $E_k/A=$200, 400, 700 and
1000 MeV moving left to right and top to bottom. The
measured points \protect\cite{bi+bi-data}
are shown by the squares.
The BUU results are for a free space cross section (solid circles),
density-dependent in-medium cross section (stars) from
(\protect\ref{Sig-med-density}) and self-consistent in-medium cross
section (open circles) from (\protect\ref{Sig-med}).
The impact parameter for all
cases is $\hat{b}\sim0.65$. There are no data
points available at 200 MeV. The curves are drawn to guide the eye.}
\label{v2-bi+bi}
\end{figure}
 In \cite{Dan-bibi} it was shown that BUU calculations of
the elliptic flow for a Bi+Bi system with a
momentum-dependent nuclear mean field favoured a
nucleon effective mass of $m^*/m\sim0.6-0.7$. We show
in figure \ref{v2-bi+bi}, the results that we obtain for BUU
calculations with the mean field in (\ref{U_single}) of effective
mass $m^*/m=0.67$,
for both a free space cross
section and separately, the in-medium cross sections obtained with 
equations
(\ref{Sig-med}) and (\ref{Sig-med-density}). The figure indicates
that adoption of the self-consistent correction to the free space
cross section is significant for the anisotropy ratio
at the lower energies
only (200 and 400 MeV/A). For the higher bombarding energies, the
self-consistent result does not differ significantly
from the free space result.
Since the in-medium
effect from (\ref{Sig-med}) probes the shape of the
momentum-dependence in the mean field it is more effective
for lower bombarding energies, with many nucleon-nucleon collisions 
occurring near the Fermi surface
where the momentum-dependence is steepest. At high bombarding
energies, the nucleon-nucleon collisions occur far from the Fermi
surface where the 
momentum-dependence in the mean field is relatively
flat, and as such, the correction to the free space cross section is
small. The experimental results are from \cite{bi+bi-data}.
Unfortunately there are no data at the lower energy, where the
differences in theoretical results are the largest. 

\begin{figure}[h!]   
\begin{center}
\includegraphics[width=6.5cm]{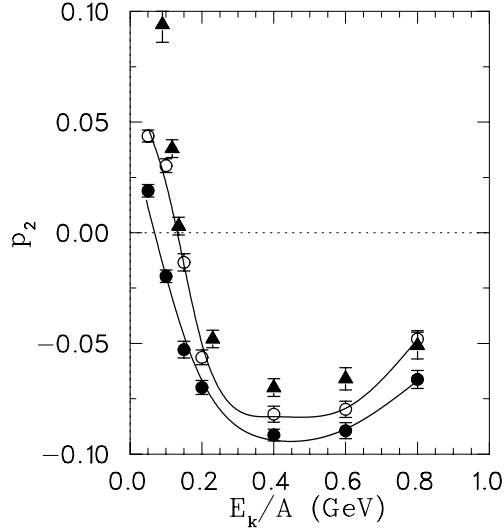}
\end{center}
\caption{Excitation function for the ellipticity coefficient for
Au+Au collisions. The measured points \protect\cite{AAndronic2} shown
by the triangles are for fragments of mass $A=1,2,3$,
and are for an impact parameter range of 
$b:5.3\rightarrow 7.3$ fm. The BUU results (circles) are as in 
figure \protect\ref{Flow-b} and are shown for an impact
parameter of $b=6.5$ fm for all mid-rapidity 
nucleons. Mid-rapidity particles are
selected with a $|y_n|<0.1$ cut.
The curves are drawn to guide the eye.}
\label{p2-E-frags}
\end{figure}
We now turn to the elliptic flow excitation function for Au+Au
recently measured by the FOPI collaboration \cite{AAndronic2}. In
figure \ref{p2-E-frags}, we show the ellipticity coefficient measured
for fragments with $A=1,2,3$ along with the appropriate BUU result.
We find that both parameterizations of the nucleon-nucleon cross section
(free space and in-medium result from (\ref{Sig-med})) are able to
qualitatively reproduce the experimental trend. Quantitatively, the calculation
with the in-medium cross-section is better able to reproduce the measured
signal.
\begin{figure}[ht!]   
\begin{center}
\includegraphics[width=6.5cm]{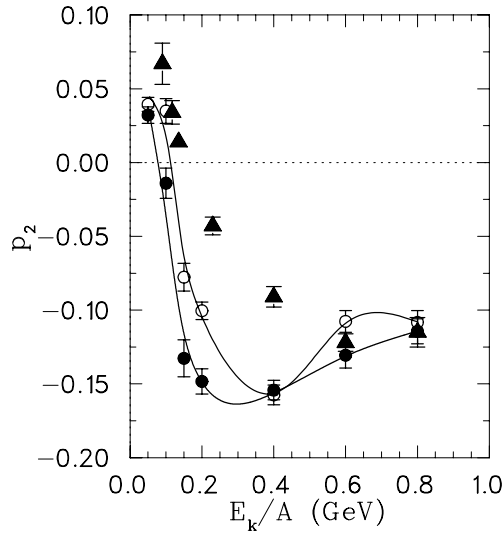}
\end{center}
\caption{Same as figure \protect\ref{p2-E-frags} but for high
transverse momentum protons
only ($p_\perp^{(0)}>0.8$). The measured points are taken from 
\protect\cite{AAndronic2} and are for an impact parameter range of 
$b:5.3\rightarrow 7.3$ fm. The BUU results (circles) as in in
figure \protect\ref{Flow-b} and are shown for free
protons at $b=6.5$ fm. Mid-rapidity particles are selected as in figure
\protect\ref{p2-E-frags}.
The curves are drawn to guide the eye.}
\label{p2-E-proton}
\end{figure}
We note that the ellipticity parameter is calculated in a frame where the
principle axis of the momentum ellipsoid
coincides with the beam axis. A more stringent test of
the model is shown in figure \ref{p2-E-proton} where the ellipticity
coefficient for free protons with transverse momenta
$p_\perp^{(0)}>$0.8 only, is plotted as a function of incident 
bombarding energy. 
Note that $p_\perp^{(0)}\equiv p_\perp/p_{beam}$.
From this figure, we find that the elliptic flow
in the energy range ($150<E_k/A<600$) MeV is over-predicted by the
model. This agrees with figure \ref{v2-bi+bi}, where we see that the
magnitude of the
elliptic flow is over-predicted at high transverse momentum.
This identifies the phase space areas where the BUU seems to perform
perhaps less satisfactorily. However there, the phase space distribution
functions are small  and therefore those regions do not contribute
importantly to the momentum-integrated observables. This is seen in our
results as well as in the experimental data. This observation however
represents valuable information for future theoretical developments.

\section{Conclusion}
In summary, we have shown for the first time the effect of the
parameter-free 
self-consistent calculation of the nucleon-nucleon in-medium
scattering cross section in BUU calculations of both directed and
elliptic flow from $E_k/A:$25$\rightarrow$800 MeV.
Our results indicate that, globally,  the in-medium
cross section better describes the flow data than with the
free space cross section. In addition, the self-consistent
calculation of the in-medium cross section results in a value of the
latter that is less than the free space cross section for low
bombarding energies and approaches the free space cross section at
high bombarding energies. While in some sense transport models utilizing
vacuum cross sections are a well-defined low-density approach, 
there is still some work to be done to
formulate and use a totally self-consistent many-body transport
theory. For example, one needs to generalize to the coupled set of transport
equations that include all additional degrees of freedom \cite{trans}.  
The work in this direction is continuing.

\acknowledgments
This work is supported in part by 
the Natural Sciences and
Engineering Research Council of Canada and in part by the Fonds FCAR of the
Qu\'{e}bec government.  One of us (D.P.) is happy to acknowledge the financial
support of McGill University through the Alexander McFee Memorial Fellowship.

\end{document}